\documentclass[prd,aps]{revtex4}

\usepackage{amsmath,amssymb,amsfonts,dcolumn,color,graphicx,graphics,latexsym,placeins,epsfig}
\usepackage{subfigure}

\newcommand{\be}{\begin{equation}}
\newcommand{\ee}{\end{equation}}
\newcommand{\ba}{\begin{eqnarray}}
\newcommand{\ea}{\end{eqnarray}}

\begin{document}

\title{\Large \bf Three dimensional Eddington--inspired
Born--Infeld gravity: solutions}

\author{Soumya Jana and Sayan Kar}
\email{soumyajana.physics@gmail.com, sayan@iitkgp.ac.in}
\affiliation{\rm Department of Physics {\it and} Center for Theoretical Studies \\Indian Institute of Technology, Kharagpur, 721302, India}

\begin{abstract}
Three dimensional Eddington-inspired Born--Infeld gravity  
is studied with the goal of finding new solutions. 
Beginning with cosmology, we obtain analytical and numerical solutions 
for the scale 
factor $a(t)$, in spatially flat ($k=0$) and spatially
curved ($k=\pm 1$) Friedmann-Roberston-Walker universes  
with (i) pressureless dust ($P=0$) and (ii) perfect fluid 
($P=\frac{\rho}{2}$), as matter sources. When the theory parameter 
$\kappa>0$,
our cosmological solutions are generically singular 
(except for the open universe, with a specific condition). 
On the other hand, for
$\kappa<0$ we do find non-singular cosmologies. 

\noindent  We then move on towards finding
static, circularly symmetric line elements with matter
obeying (i) $p=0$ and (ii) $p=\frac{\rho}{2}$. For $p=0$,
the solution found is nonsingular for $\kappa<0$ with
the matter--stress--energy representing inhomogeneous dust. 
For $p=\frac{\rho}{2}$ we obtain nonsingular solutions, for all
$\kappa$ and discuss some interesting characteristics of these
solutions. Finally, we  look at the 
rather simple $p=-\rho$ case  where the solutions
are either de Sitter or anti-de Sitter 
or flat spacetime. 

\end{abstract}


\maketitle

\section{\bf Introduction} 

\noindent Theories of gravity different from General Relativity (GR) have been 
actively pursued by many, for a variety of reasons. One such reason
relates to the possibility of avoiding the singularity problem in GR
involving the occurrence of a big bang in cosmology or black holes
in astrophysics. In a classical metric theory of gravity 
one is aware that these singularities are inevitable, as proved 
through the Hawking--Penrose theorems \cite{hawk}, under very general
and physical assumptions. However, it is quite possible that
in an alternate theory one might obtain non--singular geometries
(for example non-singular Friedman--Robertson--Walker type cosmologies)
as solutions--a feature which does not exist in the FRW cosmology based on GR.
It may be noted that the removal/resolution of a singularity 
is also expected to be a basic feature of a quantum theory of gravity.

\noindent In this article, we look into one such alternate theory. 
Historically, its origin goes back to Eddington who showed us how
de Sitter gravity could be obtained using an 
action where the Einstein--Hilbert term, $\sqrt{-g}R$ ($R$ is the 
Ricci scalar), is replaced by $\sqrt{-det (R_{ij})}$
\cite{edd}.
Eddington's formulation allowed the choice of 
the connection (instead of the metric) as
the basic variable --therefore, it is
essentially an affine formulation. However, coupling of matter
remained a problem in this formulation. 

\noindent We are also aware of Born--Infeld elecrodynamics \cite{born}, 
which was introduced in order to get rid of the infinity in the field at the
location of the charge/current. A gravity theory in the metric formulation
inspired by Born-Infeld
electrodynamics was suggested by Deser and Gibbons \cite{desgib}. 
Later Vollick \cite{vollick} worked on the various aspects of the
Deser-Gibbons proposal in a Palatini formulation 
and also introduced a non-trivial way of coupling matter in such  
theories. More recently, Banados and Ferreira  \cite{banados}
have come up with
a formulation wherein the matter coupling is different and simpler 
from that introduced in Vollick's work \cite{vollick}.  We 
will focus in the theory proposed in \cite{banados} and call it
as Eddington-inspired Born--Infeld (EiBI) gravity, for
obvious reasons. Note that the EiBI theory has the feature that it
reduces to GR, in vacuum. 

\noindent It may be noted that the theory we consider falls within the class of
bimetric theories of gravity (also called bi-gravity). The  current bimetric
theories have their origin in the seminal work of Isham, Salam and
Strathdee \cite{salam}. Numerous papers on varied aspects of 
such bimetric theories have appeared in the last few years.
The central feature here is the existence of a physical metric which
couples to matter and another auxiliary metric which is not used
for matter couplings. One needs to solve for both metrics through
the field equations.

\noindent Let us now briefly recall Eddington--inspired Born--Infeld gravity.
Since we deal with three spacetime dimensions in this article, we
prefer to write down the action and ensuing field equations in
three dimensional spacetime.
The action for the theory developed in \cite{banados}, is given as:
\begin{equation}
S_{BI}(g,\Gamma, \Psi) =\frac{2}{\kappa}\int d^3 x \left [ \sqrt{\vert g_{ij} +\kappa R_{ij}\vert}-\lambda \sqrt{-g} \right]+ S_M (g, \Psi)
\end{equation}
where $\Lambda= \frac{\lambda-1}{\kappa}$. Variation w.r.t $\Gamma (q)$, 
leads to the metric compatibility condition for the $q$ metric. Therefore,
the $q_{ij}$ is a valid Riemannian metric and is defined through 
the relation,
\begin{equation}
q_{ij}=g_{ij} + \kappa R_{ij}(q)
\label{eq:gammavarn}
\end{equation}
Variation w.r.t $g_{ij}$ gives 
\begin{equation}
\sqrt{-q} q^{ij} = \lambda \sqrt{-g}g^{ij}-\kappa \sqrt{-g} T^{ij}
\end{equation}
In order to obtain solutions, we need to assume a $g_{ij}$ and a $q_{ij}$
with unknown functions, as well as a matter stress energy ($T^{ij}$). 
Thereafter, we write down the field equations and obtain solutions
using some additional assumptions about the metric functions and the
stress energy. 

\noindent Quite some work on various fronts have been carried out
on various aspects of this theory in the last couple of years.
Astrophysical aspects have been discussed in the references in 
\cite{eibiastro} while
cosmology in those cited in \cite{eibicosmo}. Other topics such as
a domain wall brane has been analysed in \cite{eibibrane}. 
More recently, generic features of paradigms on matter-gravity couplings
have been discussed in \cite{eibigen}. However, in \cite{eibiprob}
a major problem related to surface singularities has been noticed
which has put the theory on shaky ground insofar as stellar physics
is concerned.  

\noindent Our work here is reasonably modest. In the two subsequent sections, 
we discuss cosmological and
circularly symmetric solutions in three spacetime dimensions, successively.
In the final section, we briefly summarize and conclude. 
Some of our solutions are analytical and simple. They also
maintain some of the generic features noted in the original work of Banados
and Ferreira \cite{banados}.

\section{Cosmology}
\label{sec:cosmo}
Let us assume a homogeneous and isotropic Friedmann-Robertson-Walker
(FRW) line element in 2+1 dimensions,
given as:
\begin{equation}
ds^2=-dt^2+a^{2}(t)\left[\frac{dr^2}{1-k r^2}+r^{2} d\theta^2\right]
\end{equation}
where $k=+1,0,-1$ for closed, flat and open universe respectively.
The energy--momentum tensor is taken to be that of a fluid with 
$T^{ij}=(P+\rho)u^{i}u^{j}+Pg^{ij}$. 
The conservation of energy--momentum leads to
\begin{equation}
\dot{\rho}=-2\frac{\dot{a}}{a}(P+\rho)
\label{eq:eos}
\end{equation}
which implies $\rho\propto\frac{1}{a^2}$ (for $P=0$) and $\rho\propto\frac{1}{a^3}$ (for $P=\frac{\rho}{2}$).
Further, we assume the auxiliary line element to be of the form
\begin{equation}
ds^2_{q}=-U(t)dt^2+a^{2}(t)V(t)\left[\frac{dr^2}{1-k r^2}+r^2 d\theta^2\right ]
\label{eq:auxiliarymetric}
\end{equation}
Using the auxiliary metric ($q^{ij}$), physical metric ($g^{ij}$) and 
stress-energy tensor ($T^{ij}$) in the field equation obtained by varying w.r.t.
$g_{ij}$, we get 
two equations, given by,
\begin{subequations}
\label{eq:uv}
\begin{equation}
U=\frac{D}{1+\kappa \rho_T}
\label{subeq:u}
\end{equation}
\begin{equation}
V=\frac{D}{1-\kappa P_T}
\label{subeq:v}
\end{equation}
\end{subequations}
where, $D=(1+\kappa\rho_T)(1-\kappa P_T)^2$, $\rho_T=\rho+\Lambda$, and $P_T=P-\Lambda$.   
We define two quantities $G_{2}(\rho,\Lambda)$ and $F_{2}(\rho,\Lambda)$, 
given as,
\begin{subequations}
\label{eq:gf}
\begin{equation}
G_{2}(\rho,\Lambda)=\frac{1}{\kappa}\left(1+U-\frac{2U}{V}\right)-\frac{2k}{a^2}\frac{U}{V}
\label{subeq:g}
\end{equation}
\begin{equation}
F_{2}(\rho,\Lambda)=1-\frac{\kappa(P_T+\rho_T)(1-w-\kappa P_T-\kappa w\rho_T)}{(1+
\kappa\rho_T)(1-\kappa P_T)}
\label{subeq:f}
\end{equation}
\end{subequations}
where, in Eq.~(\ref{subeq:f}), we have also assumed $P=w\rho$. 
Combining Eq.~(\ref{eq:gammavarn}) and the results from Eqs.~(\ref{eq:gf}), 
we obtain the {\em Friedmann equation}
given by,
\begin{equation}
H^2=\frac{G_2}{2F^2_2}
\label{eq:friedmann}
\end{equation}
Let us now examine two special cases: ($i$) {\em pressureless dust} ($w=0$) 
filled {\em flat} universe ($k=0$) with the cosmological constant, $\Lambda=0$
and  ($ii$) {\em radiation dominated} ($w=\frac{1}{2}$) flat universe with 
$\Lambda=0$. 

\noindent For the case $(i)$, using the equation of state 
(Eq.~(\ref{eq:eos})) and the Friedmann equation (Eq.~(\ref{eq:friedmann})), 
we find
\begin{equation}
\left( \frac{\dot{\bar{\rho}}}{\bar{\rho}} \right)^2=\frac{4\bar{\rho}(1+\bar{\rho})}{\kappa}
\label{eq:densityeqn}
\end{equation}
where, $\bar{\rho}=\kappa\rho$. Note that from the conservation law 
$\rho a^2$ is a constant. Using this in Eq.~(\ref{eq:densityeqn}), 
we obtain the solution for the scale factor $a(t)$ for $\kappa> 0$ as,
\begin{equation}
a(t)=\sqrt{C_1\left(\frac{t^2}{\kappa}-1\right)}
\label{eq:scalefactor}
\end{equation}
where the constant  $C_1=\left\lvert{\bar{\rho}}\right\rvert a^2> 0$. 
The solution clearly demonstrates that for $\kappa>0$ the EiBI theory 
cannot avoid the initial singularity ($a(t)$ can become zero at a finite
$t$ and hence we have infinite curvature). However, for $\kappa<0$, 
the Eq.~(\ref{eq:densityeqn}) becomes,
\begin{equation}
\left\lvert\frac{\dot{\bar{\rho}}}{\bar{\rho}}\right\rvert ^2=\frac{4|\rho|(1-|\rho|)}{|\kappa|}
\label{eq:densityeqn_2}
\end{equation}
From  Eq.~(\ref{eq:densityeqn_2}), we note that there exists a 
maximum density ($\rho_B$)  or, equivalently,  
a minimum value for the scale factor ($a_B$). 
The solution for the scale factor $a(t)$ for $\kappa<0$ is given by,
\begin{equation}
a(t)=\sqrt{C_2\left(\frac{t^2}{|\kappa|}+1 \right)}
\label{eq:scalefactor_2}
\end{equation}
where $C_2$ is a constant.
It is easy to see that the scale factor is never zero and thus, there is no
curvature singularity. Both these solutions are plotted in the top row of
Fig. 1.

\noindent For case ($ii$),(i.e radiation dominated universe), 
the Friedmann equation becomes:
\be
H^2=\frac{2}{\kappa}\left[ 2\bar{\rho}-1+\left(1-\frac{\bar{\rho}}{2}\right)^2(1+\bar{\rho})-\frac{\kappa k}{a^2}(2-\bar{\rho})\right]\frac{(1+\bar{\rho})(2-\bar{\rho})^2}{(4-\bar{\rho}+4\bar{\rho}^2)^2}
\label{eq:friedmann_rad}
\ee

\begin{figure}[h]
\begin{center}
\mbox{\epsfig{file=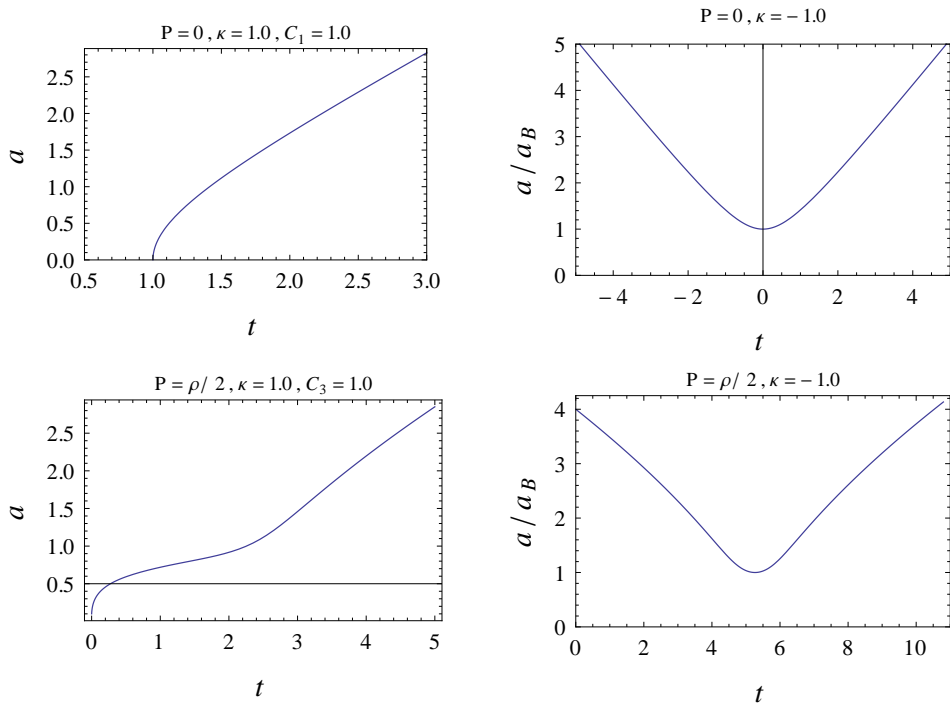,width=5in,angle=360}}
\end{center}
\caption{Plot of the time evolution of the scale-factor $a(t)$ for flat universe ($k=0$).The parameters are specified
on each frame. $a_B$ is the minimum scale factor for non-singular universe. $C_1$ and $C_3$ are constants as defined in the text.}
\label{fig:cosmology}
\end{figure}

For the {\em flat universe} ($k=0$), the last term in the square-bracket of the right hand side of the Eq.~(\ref{eq:friedmann_rad}) does not contribute and the equation becomes,
\begin{equation}
H^2=\frac{2}{k}\left[ 2\bar{\rho}-1+\left(1-\frac{\bar{\rho}}{2}\right)^2(1+\bar{\rho})\right]\frac{(1+\bar{\rho})(2-\bar{\rho})^2}{(4-\bar{\rho}+4\bar{\rho}^2)^2}
\label{eq:friedmann_1}
\end{equation}
In this case, for $\kappa>0$, $H^2\geq 0$ for an arbitrary, but physically 
justifiable value of $\bar{\rho}$ (i.e for $\bar{\rho}\geq0$). 
Thus, here also, a curvature singularity appears in the solution.
Using the equation of state :$\bar{\rho}a^3=C_3$, where $C_3$ is a constant, we can solve numerically the Eq.~(\ref{eq:friedmann_1}) for the time evolution 
of the scale-factor $a(t)$, which is shown in Fig.~\ref{fig:cosmology}. 
However, for $\kappa<0$, the Eq.~(\ref{eq:friedmann_1}) is rewritten as,
\begin{equation}
H^2=\frac{2}{|\kappa|}\left[ 1+2|\bar{\rho}|+\left( 1+\frac{|\bar{\rho}|}{2}\right)^2(|\bar{\rho}|-1)\right]\frac{(1-|\bar{\rho}|)(2+|\bar{\rho}|)^2}{(4+|\bar{\rho}|+4|\bar{\rho}|^2)^2}
\label{eq:friedmann_2}
\end{equation}
The presence of the factor $(1-|\bar{\rho}|)$ in the Eq.~(\ref{eq:friedmann_2}) leads to $H^2$ being negative when $|\bar{\rho}|>1$ . Therefore, 
in this case, for $\kappa<0$, there exists a maximum density or, 
equivalently, a minimum, non--zero value for the scale factor. 

\noindent Let us now define two dimensionless variables 
$\Omega=\frac{\rho}{\rho_B}$ ($0\leq\Omega\leq 1$),$z=\frac{a}{a_B}$. Using 
these we recast the equations as:
\begin{subequations}
\label{eq:friedmann_3}
\begin{equation}
|\kappa H^2(\Omega)|=2\left[ 1+2\Omega+\left(1+\frac{\Omega}{2}\right)^2(\Omega-1)\right]\frac{(1-\Omega)(2+\Omega)^2}{(4+\Omega+4\Omega^2)^2}
\label{subeq:friedmann_3a}
\end{equation}
\begin{equation}
\dot{z}^2=\frac{2z^2}{|\kappa|}\left[ 1+\frac{2}{z^3}+\left(1+\frac{1}{2z^3}\right)^2\left(\frac{1}{z^3}-1\right)\right]\frac{\left(1-\frac{1}{z^3}\right)\left(2+\frac{1}{z^3}\right)^2}{\left(4+\frac{1}{z^3}+\frac{4}{z^6}\right)^2}
\label{subeq:friedmann_3b}
\end{equation}
\end{subequations}
Numerically solving the above equation, we plot and note the 
time evolution of the scale factor using the Eqs.~(\ref{eq:friedmann_3}). 
In the bottom row of Fig.~\ref{fig:cosmology}
we note a non-singular scale factor for $\kappa<0$ --a  result
showing the existence of a bounce, similar to that obtained 
in $3+1$ dimensions. For $\kappa>0$, the solution appears to be singular.
The above solutions (especially, the ones for the $P=0$ case) are instructive 
because they are, as far as we know, the only 
known analytical solutions in EiBI cosmology. 

Introducing curvature in the spatial slices (i.e. $k=+1,-1$) does not yield
anything drastically new in the dust ($P=0$) case primarily because of
the fact that $\rho\propto \frac{1}{a^2}$.The solutions for $\kappa>0$ and
$\kappa<0$ are given by,
\begin{subequations}
\label{eq:arad}
\begin{equation}
a(t)=\sqrt{\left(1-\frac{k}{\rho_0}\right)C_1\left(\frac{t^2}{\kappa}-1\right)} ,\kappa>0
\label{subeq:arad_a} 
\end{equation}
\begin{equation}
a(t)=\sqrt{\left(1-\frac{k}{\rho_0}\right)C_2\left(\frac{t^2}{|\kappa|}+1\right)} , \kappa<0
\label{subeq:arad_b}
\end{equation}
\end{subequations}
where $\rho_0$ is a constant ($\rho a^2=\rho_0 $ for $P=0$). 
For a $k=+1$ solution there is a lower bound on $\rho_0$ ($\rho_0>1$) 
whereas, for $k=-1$, $\rho_0$ is an arbitrary positive, real constant.
When $k=0$, we recover the earlier results (Eq.~\ref{eq:scalefactor} and Eq.~\ref{eq:scalefactor_2} ). It is easy to see that there 
is nothing new in the $P=0,k\neq 0$ solutions. 

\begin{figure}[h]
\begin{center}
\mbox{\epsfig{file=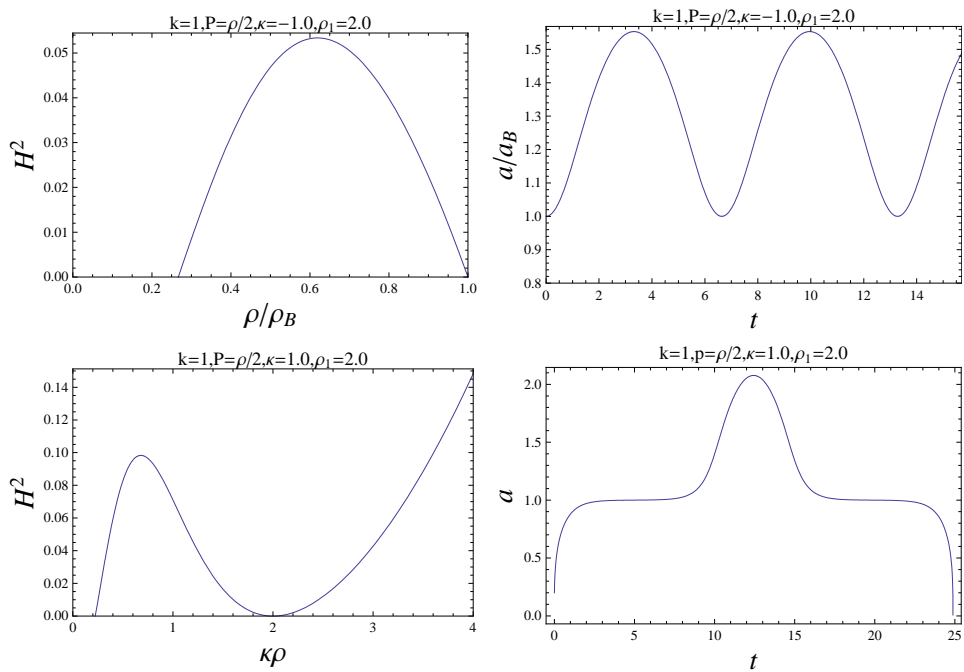,width=5in,angle=360}}
\end{center}
\caption{Plot of $H^2$ and the time evolution of the scale-factor $a(t)$ for 
closed universe ($k=+1$).
$\rho_{1}$ is a constant,($\rho a^3=\rho_{1}$).}
\label{closeduniv}
\end{figure}

\begin{figure}[h]
\begin{center}
\mbox{\epsfig{file=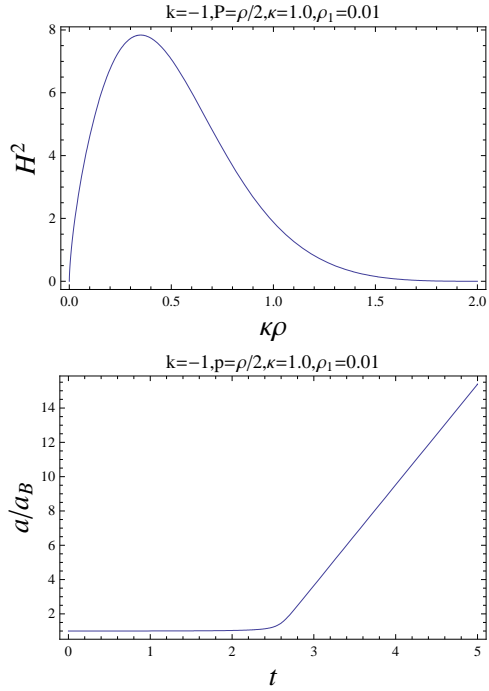,width=2.5in,angle=360}}
\end{center}
\caption{Plot of $H^2$ and $a(t)$ for open universe ($k=-1$) showing the 
{\em loitering phase} at its early stage.
$\rho_{1}$ is a constant,($\rho a^3=\rho_{1}$).}
\label{openuniv}
\end{figure}

However, in the radiation dominated ($P=\frac{\rho}{2}$) case we do get some 
interesting results, though the main conclusion regarding the appearance of 
a singularity or otherwise, is almost the same. The introduction of
spatial curvature results in an additional term as shown in the square-bracket 
of  the R.H.S. of Eq.~(\ref{eq:friedmann_rad}) and this leads to all the 
differences. For a {\em closed universe} ($k=+1$), instead of a
{\em bounce} we get an {\em oscillation} 
of the universe for $\kappa<0$. For $\kappa>0$ there is an additional 
feature implying a maximum value of the scale-factor, 
along with the singularity. These are shown in Fig.\ref{closeduniv}.
In  an {\em open universe} ($k=-1$), we do not see any characteristic 
novelties for $\kappa<0$, but for $\kappa>0$
along with the singularity, we also get, under certain circumstances, 
a non-singular {\em loitering} phase of the early universe. This is shown 
through the plots in Fig.~\ref{openuniv}.

The main differences, in the context of cosmology, 
between the results in $2+1$ dimensions and 
those in the $3+1$ dimensional version of EiBI theory 
are the following.
In $2+1$ dimensions, for $\kappa>0$, the cosmological solutions
are singular for  matter satisfying $P=0$ and $P=\rho/2$. In contrast,
the $3+1$ dimensional cosmological solutions are non-singular for both 
$P=0$ and $P=\frac{\rho}{3}$.
Further, in $2+1$, for $P=0$, we find a bounce solution
for $\kappa<0$, whereas in $3+1$, as obtained by I. Cho {\em et. al.}\cite{eibicosmo}(see Fig.2 there), 
the early universe is de Sitter spacetime and $H^2$ is constant at early times. 
In \cite{eibicosmo}, an analytical expression for
the scale factor $a(t)$ in the $3+1$ dimensional $P=0$ case,
has indeed been obtained 
but the expression involves non-invertible functions. In $2+1$, with
$P=0$, we find analytical scale factors which involve simple functions. 
The loitering phase of the early universe for $\kappa>0$ in $3+1$ dimension is 
absent in $2+1$, except for a specific situation
with an open universe ($k=-1$). For a radiation dominated closed universe 
($k=+1$), we obtain a big crunch solution for $\kappa>0$ 
(see Fig.~\ref{closeduniv}),
whereas, in $3+1$ dimensions, the solution under a similar situation 
has a non-singular loitering phase. Thus, the lower dimensional
toy model cosmologies have expected similarities and
differences when compared with their
higher dimensional counterparts.

\section{Circularly symmetric, static solutions}
\label{sec:spherically symmetric}
Let us now turn to a completely different class of line elements
--i.e. those which are circularly symmetric and static. 
We consider a simple ansatz for the physical line element $g_{ij}$,
\begin{equation}
ds^2=-f^2(l)dt^2+dl^2+r^2(l)d\theta^2
\label{eq:sph_orgnl}
\end{equation}
where $r(l)$, $f(l)$ are non--negative functions and $l$ extends from 
minus infinity to plus infinity. $r(l)$ represents the radius of a circle at each value of $l$. $f^2(l)$ is the so-called redshift function.
The energy-momentum tensor is assumed as, $T^{ij}$=diag.$\left(\frac{\rho}{f^2(l)},p_1,\frac{p_2}{r^2(l)} \right)$. Let us further assume the auxiliary line element
to be of the form
\begin{equation}
ds^2_q=-h^2(l)dt^2+u^2(l) \left [ dl^2+ r^2(l)d\theta^2\right ]
\label{eq:sph_aux}
\end{equation}
where $u(l)$ and $h(l)$ are non--negative functions of $l$. 
The field equation obtained from $g_{ij}$ variation
yields,
\begin{equation}
\rho= \frac{1}{\kappa} \left (\frac{f}{h} u^2-1\right ) -\Lambda
\hspace{0.2in} ;\hspace{0.2in} p_1=p_2=p=\Lambda +\frac{1}{\kappa}\left (1-\frac{h}{f}\right )
\label{eq:prhodef}
\end{equation}
The other field equation obtained from $\Gamma$ variation yield the
following equations,
\begin{eqnarray}
1-\frac{f^2}{h^2} = -\frac{\kappa}{u^2}\left ( \frac{h''}{h} +\frac{h'}{h}\frac{r'}{r}\right )\label{eq:cirgamma1} \\
u^2-1=-\kappa\left[\frac{h''}{h} - \frac{h'}{h}\frac{u'}{u} + \frac{r''}{r}+\frac{r'u'}{ru}+\frac{u''}{u}-\left(\frac{u'}{u}\right)^2\right]
\label{eq:cirgamma2}\\
u^2-1=-\kappa\left[\frac{h'}{h}\frac{r'}{r} + \frac{h'}{h}\frac{u'}{u} + \frac{r''}{r}+\frac{r'u'}{ru}+\frac{u''}{u}-\left(\frac{u'}{u}\right)^2\right]
\label{eq:cirgamma3}
\end{eqnarray}
Consistency of the last two equations (since the L. H. S. of both are the same)
leads to the simple relation
\begin{equation}
h'=\pm C_1 u^2 r
\label{eq:consistency}
\end{equation}
where $C_1$ is a constant. Further, the conservation law $\nabla_j T^{ij}=0$
implies
\begin{equation}
p'+\frac{f'}{f} \left (\rho+p\right ) =0
\label{eq:eos1}
\end{equation}
Using the expressions for $\rho$ and $p$ given earlier, we arrive at 
\begin{equation}
h h' =f f' u^2
\end{equation}
which, using the expression $h'=\pm C_1 u^2 r$ becomes
\begin{equation}
ff' = \pm C_1 r h
\label{eq:eos2}
\end{equation}

We can now look for possible solutions of the above equations. 

\subsection{ $C_1=0$, $p=0$}

The first of these involves
assuming $C_1=0$ which, from the equations imply $p=0$. 
Equivalently, $f=h=1$, i.e. $f$ and $h$ are
both constants. With $f=h=1$, Eqn (\ref{eq:cirgamma1}) is vacuous and the 
remaining single equation for $u^2-1$ is
\begin{equation}
u^2-1=-\kappa\left[\frac{r''}{r}+\frac{r'u'}{ru}+\frac{u''}{u}-\left(\frac{u'}{u}\right)^2\right]
\label{eq:sph_gm2}
\end{equation}
To solve this equation we need another condition. With $h=f=1$, we note
that $p=p_1=p_2=\Lambda$ and hence there is no scope of assuming an
equation of state. In other words the solution we are looking for
is produced by {\em inhomogeneous dust} in the presence of a $\Lambda$.
To find one such solution, we assume a relationship
between $u$ and $r$. Let us take
\begin{equation}
\frac{r'}{r}+\frac{u'}{u}=D
\label{eq:sph3}
\end{equation}
where $D$ is a constant.
Thus, we can transform the second order differential 
equation (Eq.~\ref{eq:sph_gm2}) into an easily solvable first order ordinary
differential equation for $u$. Other choices for the R. H. S. of 
Eq.~\ref{eq:sph3}
(may not be a constant)  could give other
mathematical solutions for which we need to 
verify whether the energy-density distributions ($\rho_T$) are physically
acceptable. 

\begin{figure}[h]
\begin{center}
\mbox{\epsfig{file=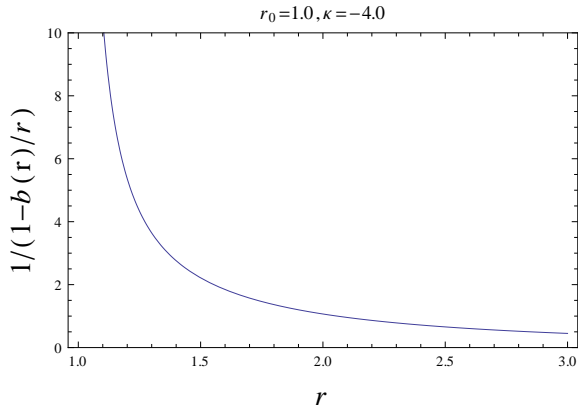,width=3in,angle=360}}
\end{center}
\caption{Plot of $\frac{1}{1-\frac{b(r)}{r}}$ }
\label{fig:sph_b(r)}
\end{figure}

\noindent We now demonstrate 
one solution assuming the condition in Eq.~(\ref{eq:sph3}) and 
$D=1/\sqrt{\vert\kappa\vert}$.
Let us also assume $\kappa<0$.
The solution we obtain is specified by the $r(l)$ and $u(l)$ given below:
\begin{eqnarray}
r^2=r_0^2 \cosh \frac{2l}{\sqrt{\vert\kappa\vert}} \hspace{0.2in};\hspace{0.2in}
u=\frac{e^{\frac{l}{\sqrt{\vert \kappa\vert}}}}{\sqrt{\cosh{\frac{2l}{\sqrt{\vert
\kappa\vert}}}}}
\end{eqnarray}
The energy density is given as 
\begin{equation}
\rho= -\frac{1}{\vert\kappa\vert}\tanh \frac{2l}{\sqrt{\vert\kappa\vert}} - \Lambda
\end{equation}
and is positive as long as $\Lambda<0$ and 
$\frac{1}{\sqrt{\vert \kappa\vert}} < \vert \Lambda \vert$.
The corresponding Ricci scalar for the physical $g$ metric is,
\begin{equation}
R=-\frac{2+2sech^2\left(\frac{2l}{\sqrt{|\kappa|}}\right)}{|\kappa|}
\end{equation}
which is clearly non-singular and has asymptotically constant 
negative curvature. 
The {\em Kretschmann scalar} is
\begin{equation}
 K=R^{ijkl}R_{ijkl}=\frac{4}{|\kappa|^2}\left[2-\tanh^2\left(\frac{2l}{\sqrt{|\kappa|}}\right)\right]^2
\end{equation}
and is non-singular as well. The
physical line element, for this case   
can be written as,
\begin{equation}
ds^2=-dt^2+dl^2+r^2_0\cosh\left(\frac{2l}{\sqrt{|\kappa|}}\right)d\theta^2
\label{originalmetric}
\end{equation}
One can also rewrite it as,
\begin{equation}
ds^2=-dt^2+\frac{dr^2}{1-\frac{b(r)}{r}}+r^2d\theta^2
\label{modf_originalmetric}
\end{equation}
where, 
\begin{equation}
b(r)=r+\frac{1}{|\kappa|}\left(\frac{r_0^4}{r}-r^3\right)
\end{equation}
and $r=r_0\sqrt{\cosh\left(\frac{2l}{\sqrt{|\kappa|}}\right)}$. 
Thus, $r\geq r_0$ and a metric
singularity occurs in the transformed metric function only at $r=r_0$. 
We plot the transformed metric function $1/(1-\frac{b(r)}{r})$ and have 
shown it in Fig.~\ref{fig:sph_b(r)}.
Note that this solution is not asymptotically flat (i.e. $\frac{b(r)}{r}$ does
not tend to zero as $r\rightarrow\infty$). However, there is a minimum
radius $r=r_0$ and the geometry is symmetric as $l\rightarrow-l$. Also
$b(r=r_0)=r_0$. Thus, the features are similar to that of a 
Lorentzian wormhole \cite{wormhole} though the geometry is 
not asymptotically flat (in fact it is asymptotically anti-de Sitter).  

\begin{figure}[ht]
\centering
\subfigure[]{
\mbox{\epsfig{file=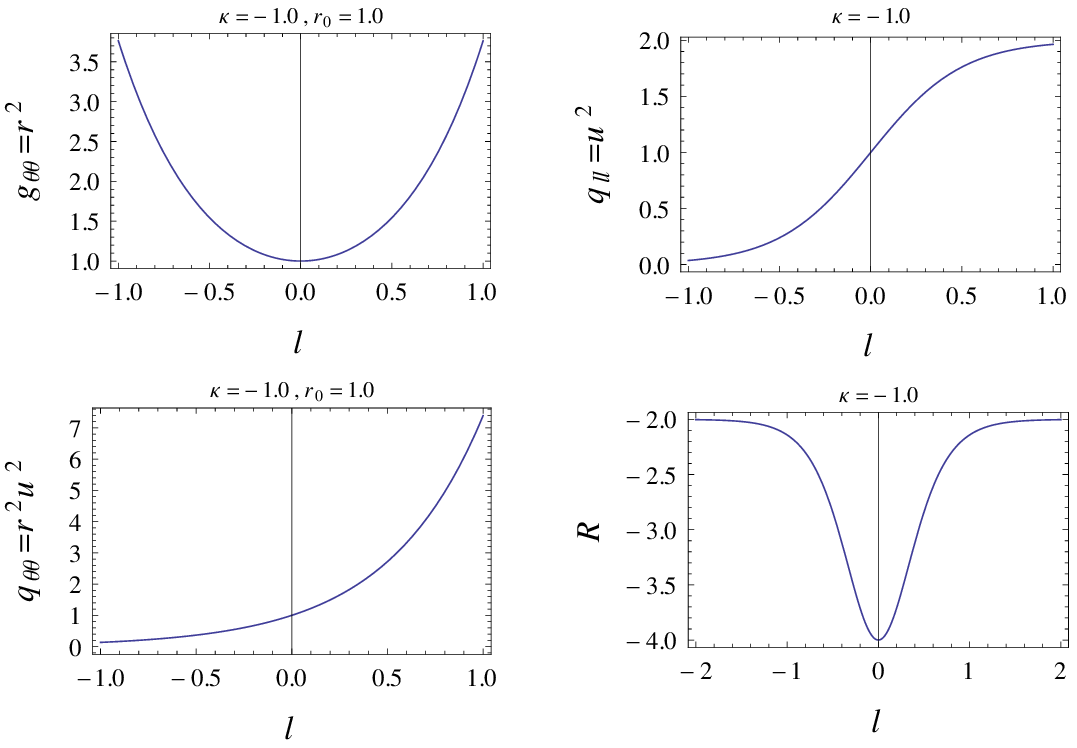,width=6.0in,angle=360}}
\label{fig:dustgeo}}
\subfigure[]
{
\mbox{\epsfig{file=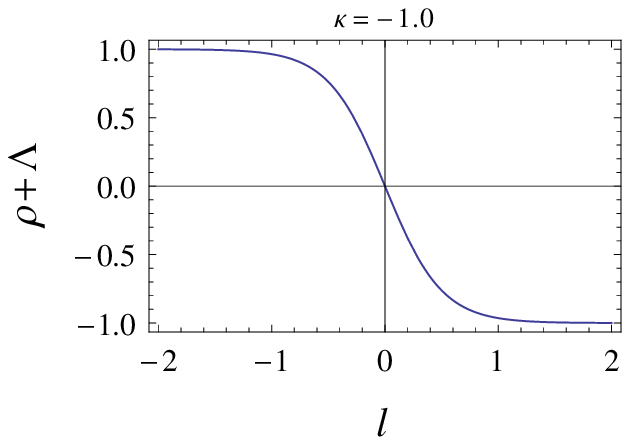,width=3.0in,angle=360}}
\label{fig:dustrho}
}
\caption{
(a)Plot of physical metric function $g_{\theta\theta}$, auxiliary metric functions $q_{ll}$ and $q_{\theta\theta}$ and the Ricci scalar ($R$) for $p=\Lambda$.
(b)Plot of energy density($\rho_T=\rho+\Lambda$). The value of the parameters ($r_0, \kappa$) are shown on the frame of each plot.}
\label{fig:dustsoln}
\end{figure}

\noindent In Fig.~\ref{fig:dustsoln} we have shown the $g$ and $q$
metric functions, the Ricci scalar ($R$), and the energy-density ($\rho_T=\rho+\Lambda$)  for the solution quoted above. 

\noindent We note that the energy-density is asymmetric as $l\rightarrow -l$,
even though the physical metric $g$ is symmetric in $l$. This happens 
because the $q$ metric is asymmetric and the field equation which
contains the energy--momentum tensor, has contributions from the $g$
and $q$ metrics. The asymmetry is also evident from the expression for
$\rho$ which clearly depends only on $u^2$. 

\noindent If $\kappa>0$, assuming the same relation (Eq.~\ref{eq:sph3}) between $r$ and $u$, we get a singular solution.
In this case, $l$ becomes restricted and at the boundary of $l$, both the Ricci scalar and the energy density ($\rho$)
diverge.

\subsection{$C_1\neq 0$, $p=\frac{\rho}{2}$}

In this case, the equations we need to solve are Eqns. (\ref{eq:prhodef}), (\ref{eq:cirgamma1}), (\ref{eq:cirgamma2}) or
(\ref{eq:cirgamma3}), (\ref{eq:consistency}) and (\ref{eq:eos2}) which are six equations in the six variables
$\rho$, $p$, $f$, $h$, $r$ and $u$. However, we now show that
Eqns (\ref{eq:cirgamma1}) and (\ref{eq:cirgamma3}) (or (\ref{eq:cirgamma2})) are not independent equations. 
To this end, we define
\begin{equation}
\eta= \frac{r'}{r} \hspace{0.2in};\hspace{0.2in} \xi=\frac{u'}{u}
\hspace{0.2in};\hspace{0.2in} \chi=\xi+\eta
\end{equation} 
Using these definitions and the other equations, one can show that
Eqn. (\ref{eq:cirgamma1}) reduces to
\begin{equation}
\chi = \pm \left(\frac{f^2}{2 \kappa C_1 r h} - \frac{h}{2 C_1 r \kappa}\right)
\end{equation}
Further, one may rewrite Eqn (\ref{eq:cirgamma3}) using the new variables defined
above. $\chi'$ as obtained from the reduced version of (\ref{eq:cirgamma1})
matches with the $\chi'$ as obtained from (\ref{eq:cirgamma3}). Hence these
two equations are not independent. We exploit this feature of
the system of equations 
to impose an equation of state restriction. This is chosen to
be $p=\frac{\rho}{2}$. 

\begin{figure}[h]
\begin{center}
\mbox{\epsfig{file=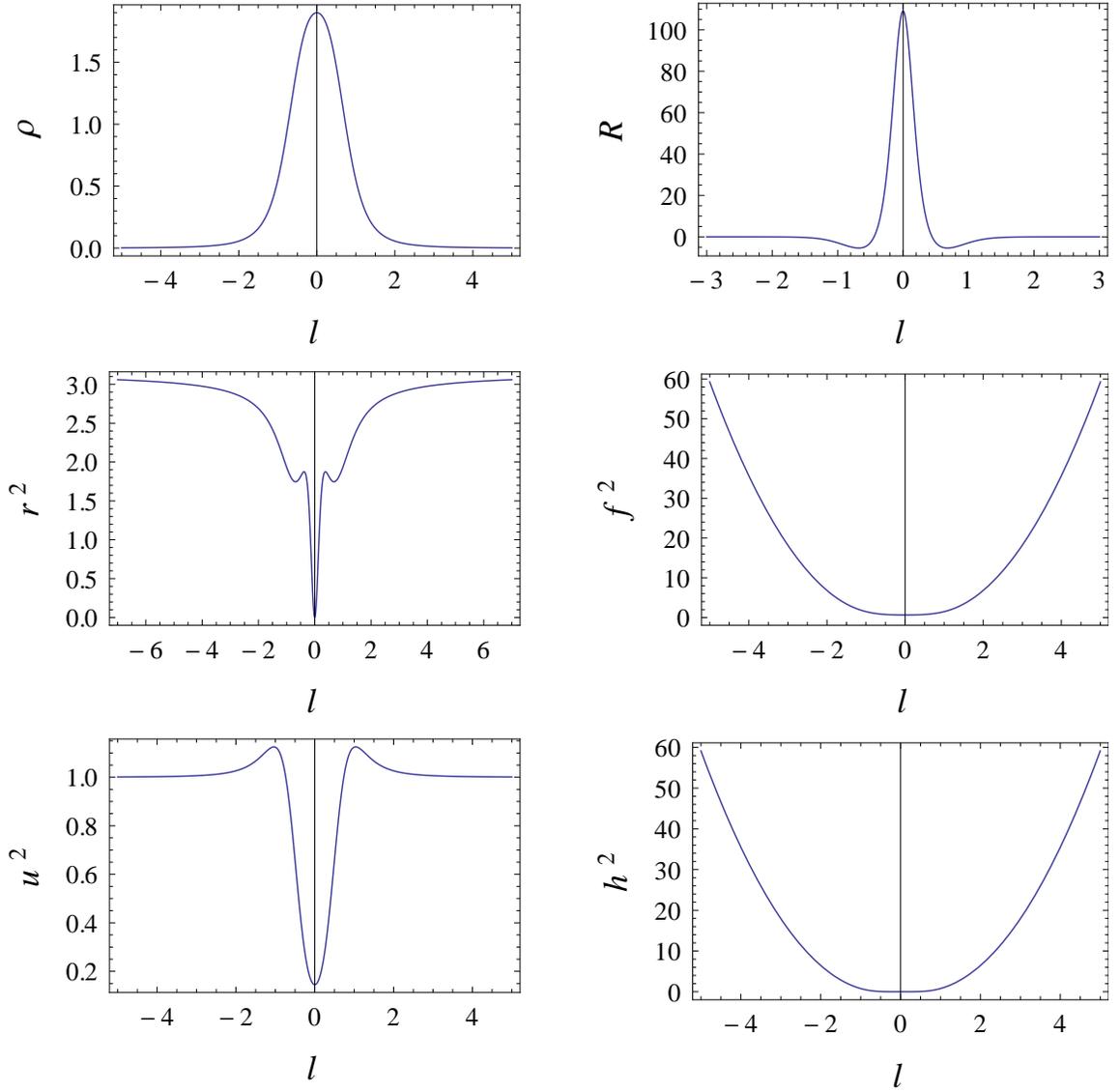,width=6.0in,angle=360}}
\end{center}
\caption{Plot of all metric functions, energy density and Ricci scalar for $p=\rho/2$, using $\kappa=1$,$C_1=C_4=1$,$\rho(0)=1.9$ and $\rho'(0)=0$.}
\label{fig:rad1}
\end{figure}

\begin{figure}[h]
\begin{center}
\mbox{\epsfig{file=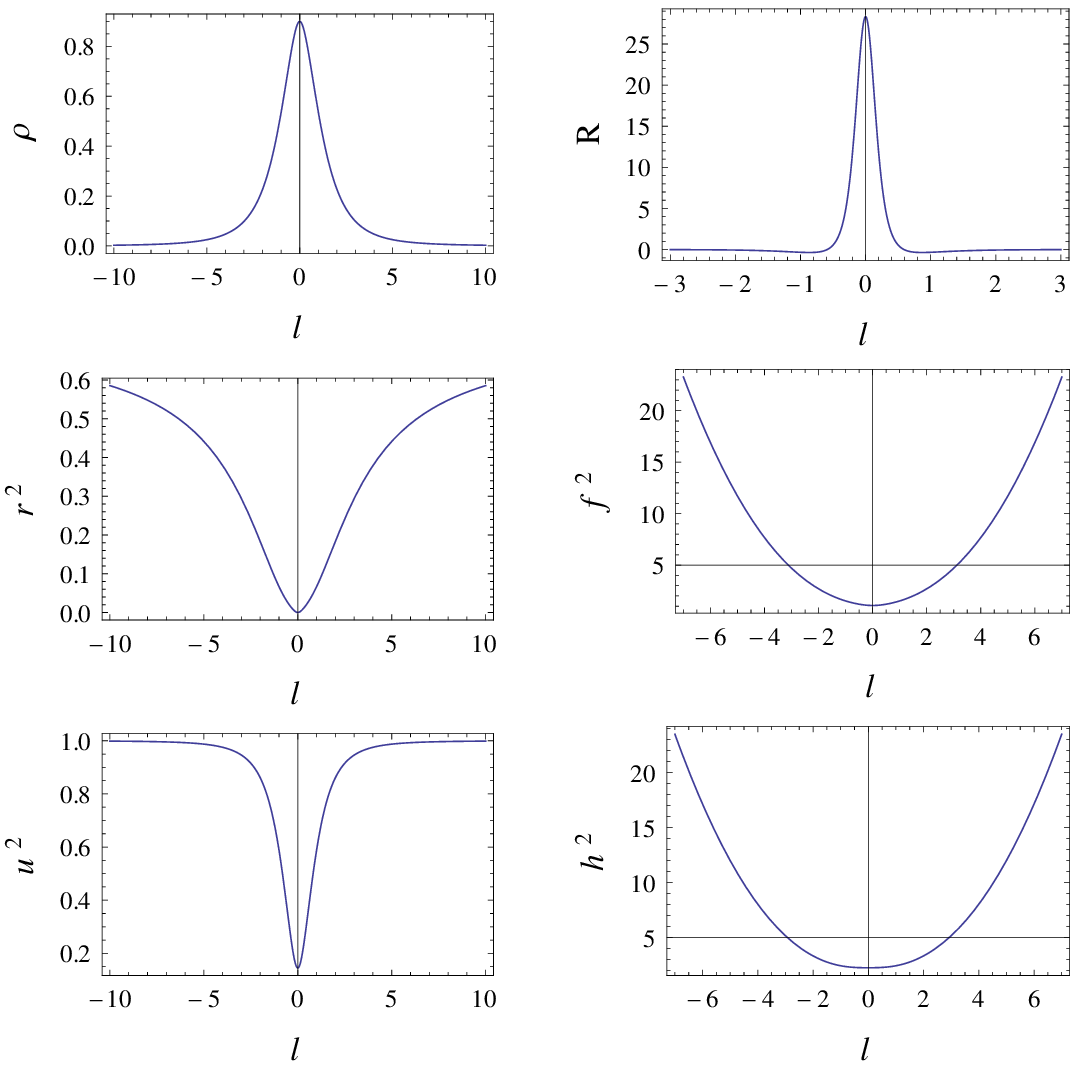,width=6.0in,angle=360}}
\end{center}
\caption{Plot of all metric functions, energy density and Ricci scalar for $p=\rho/2$, using $\kappa=-1$,$C_1=C_4=1$,$\rho(0)=0.9$ and $\rho'(0)=0$.}
\label{fig:rad2}
\end{figure}

\noindent Using the above equation of state, it is straightforward to write down all
the unknown functions as functions of $\rho$. These are given below as,
\begin{eqnarray}
u^2 = \frac{1}{2} \left (1+\bar\rho\right )\left (2-\bar\rho\right ) \label{eq:urho}\\
f= C_4 \left (\frac{\kappa}{\bar \rho}\right )^{\frac{1}{3}} \label{eq:frho}\\
h= \frac{1}{2} \left (2-\bar \rho\right ) C_4 \left (\frac{\kappa}{\bar \rho}\right )^{\frac{1}{3}} \label{eq:hrho}\\
r=\pm \frac{2}{3}  \frac{C_4}{C_1} \left (\frac{\kappa}{\bar \rho}\right )^{\frac{1}{3}} \frac{\bar \rho'}{\bar \rho(2-\bar \rho)}
\label{eq:rrho} 
\end{eqnarray}
where, we have assumed $\Lambda=0$, defined $\bar\rho=\kappa\rho$ and 
introduced a non-zero positive constant $C_4$. 
Further, we have, from the field equations, the following equation for 
$\bar \rho$,
\begin{equation}
\bar\rho''+\frac{1}{6} \frac{\bar\rho'^2 \left (8\bar\rho^2 + \bar \rho-16\right )}{\bar\rho (1+ \bar \rho )(2-\bar\rho)} +\frac{3}{8\kappa} \bar\rho^2
(4-\bar\rho) =0
\label{eq:rho}
\end{equation}
Due to the presence of the factor $(1+\bar\rho)(2-\bar\rho)$ in the denominator of the second term in the L.H.S, the Eqn.~(\ref{eq:rho})
can be solved only for $\bar\rho_{max}<2$ for $\kappa>0$ and $\vert\bar\rho_{max}\vert<1$ for $\kappa<0$. 
Though the range of $\rho$ is restricted but $\rho_{max}$ can approach 
its limiting values and be as close as required.
One can obtain qualitative analytical information about the nature
of the solutions by using approximations. For instance, when $\kappa>0$,
we can show, using a Taylor series about $\bar\rho\rightarrow 2$, that
the solution will behave as $-\frac{3}{2} l^2 + const$. On the other hand,
when $\bar\rho\rightarrow 0$, the solution behaves generically as 
$\frac{1}{(a l +b)^3}$ ($a$, $b$ being constants). 
The equation for $\bar \rho$ can be easily solved numerically and the
solutions for $\kappa>0$ and $\kappa<0$ are shown in Figs~\ref{fig:rad1},\ref{fig:rad2}.  
Note that the plot for $\bar\rho$ as a function of $l$ matches
qualitatively with the approximate solutions found in the neighborhoods of
$\bar\rho=2$ and $\bar\rho=0$. 

We also note that the Ricci scalar is regular everywhere though the
$r(l)$ does become zero at $l=0$.
We will now try to see analytically, why the 
solution is regular everywhere. The expression for the Ricci
scalar as a function of $\bar\rho$ turns out to be
\begin{equation}
R=-2 \left[\frac{\bar\rho'''}{\bar\rho'}+\bar\rho''\left(\frac{3}{2-\bar\rho}-\frac{14}{3\bar\rho}\right)+\bar\rho'^{2}\left(\frac{4}{\bar\rho^2}+
 \frac{2}{(2-\bar\rho)^2}-\frac{3}{\bar\rho(2-\bar\rho)}\right)\right]
\label{eq:radricci}
\end{equation}
From the differential equation for $\bar\rho$ we can show, by taking
another derivative w.r.t. $l$, that $\frac{\bar\rho'''}{\bar\rho'}$ is
finite everywhere including the location of the maximum density
at $l=0$ where $\bar\rho'=0$ and $\bar\rho''<0$. The remaining terms 
are all finite everywhere and thus the Ricci scalar is finite.
One can also check that the Kretschmann scalar
\begin{equation}
R_{ijkl}R^{ijkl}=4\left[\left(\frac{f''}{f}\right)^2+\left(\frac{r'f'}{rf}\right)^2+\left(\frac{r''}{r}\right)^2\right]
\label{eq:radk}
\end{equation}
is  finite everywhere and therefore $R_{ij}R^{ij}$ is also
finite. Thus at the location of a maximum density the geometry
remains nonsingular though the radius $r(l)$ becomes zero. 
This implies the vanishing of the circumference of the
circle at $t=$constant and $l=0$. The vanishing of $r(l)$ at $l=0$
would have implied a singularity (as in cosmology, where $a(t)=0$ implies
a singularity) in GR.
However, in EIBI theory it does not
imply a curvature singularity essentially due to presence of a non--zero
$\kappa$. In the cosmological context,
such regular solutions with vanishing volume spatial hypersurfaces
have been noted earlier in the literature \cite{van}. 
The form of $f^2(l)$ as shown in Figs.~\ref{fig:rad1},\ref{fig:rad2} 
indicates the
absence of any horizon ($f(l)$ is never zero) in the geometry. 
The energy density and the 
pressure are both finite and positive definite everywhere as is clear
from the graphs in Figs.~\ref{fig:rad1},\ref{fig:rad2}. We also note that 
the spacetime is asymptotically flat and tends to one of the 
vacuum ($p=0,\rho=0,\Lambda=0$) solutions,  $r=$const., $u^2=1$ and $h^2=f^2=(cl+d)^2$ ($c$ and $d$ being constants), at infinity.

\subsection{ $C_1\neq 0$, $p=-\rho$}
In this simple case we solve a similar set of five equations, 
 assuming the equation of state $p=-\rho$, with $\Lambda=0$. 
From the Eqn.~(\ref{eq:eos1}), we have $p=-\rho_0$ ($\rho_0$ is a positive constant). 
Using this in Eqn.~(\ref{eq:prhodef}), we get $u^2=(1+\bar\rho_0)^2$
and $\frac{f^2}{h^2}=\frac{1}{(1+\bar\rho_0)^2}$. 
Further, using these results in the remaining equations, we finally obtain 
the second order 
linear homogeneous differential equation for $h$, which is 
\begin{equation} 
h''+\frac{\bar\rho_0(\bar\rho_0+2)}{2\kappa}h=0
\label{eq:h}
\end{equation}
Solving the Eqn.~(\ref{eq:h}) and using other relations, we get trivial but non-singular solutions. 
For $\kappa>0$, the solutions for $h(l)$,$f(l)$ and $r(l)$ are
\begin{eqnarray*}
h=\left\{\sin\left(\sqrt{\frac{\bar\rho_0(\bar\rho_0+2)}{2\kappa}}l\right),\cos\left(\sqrt{\frac{\bar\rho_0(\bar\rho_0+2)}{2\kappa}}l\right)\right\}
=f(1+\bar\rho_0)\\
r=\frac{\sqrt{\bar\rho_0(\bar\rho_0+2)/2\kappa}}{C_1(1+\bar\rho_0)^2}\left\{\cos\left(\sqrt{\frac{\bar\rho_0(\bar\rho_0+2)}{2\kappa}}l\right)
,\sin\left(\sqrt{\frac{\bar\rho_0(\bar\rho_0+2)}{2\kappa}}l\right)\right\}
\end{eqnarray*}
The resulting physical line element is de-Sitter spacetime. Similarly, 
for $\kappa<0$, we get de-Sitter spacetime (if $\vert\bar\rho_0\vert<2$ but $\vert\bar\rho_0\vert\neq 1$).
For $\kappa<0$ and $\vert\bar\rho_0\vert>2$, we get,
\begin{eqnarray*}
h=\exp\left(\pm\sqrt{\frac{\vert\bar\rho_0\vert(\vert\bar\rho_0\vert-2)}{2\vert\kappa\vert}}l\right)=f(\vert\bar\rho_0\vert-1)\\
r=\frac{\sqrt{\vert\bar\rho_0\vert(\vert\bar\rho_0\vert-2)/2\vert\kappa\vert}}
{C_1(1-\vert\bar\rho_0\vert)^2}\exp\left(\pm\sqrt{\frac{\vert\bar\rho_0\vert(\vert\bar\rho_0\vert-2)}{2\vert\kappa\vert}}l\right)
\end{eqnarray*} 
This solution represents  anti-de Sitter spacetime. 
For $\kappa<0$ and $\vert\bar\rho_0\vert=2$, we find,
\begin{eqnarray*}
h=Al+B=f \hspace{0.1in},\mbox{(A,B being constants)}, \\
r^2=\frac{A^2}{C_1^2}, (\mbox{constant})
 \end{eqnarray*}
which is just flat spacetime.
 
\section{Conclusions}
The main purpose behind this article was to find simple solutions in
three dimensional EiBI gravity. We have focused here on two types of solutions
--cosmological and static, circularly symmetric. We summarise our results 
pointwise below.

$\bullet$ We have found 
analytical, spatially flat cosmologies for pressureless dust ($P=0$) and 
numerical solutions for $P=\frac{\rho}{2}$. We have also explored
the cases when spatial curvature is present.
The cosmological solutions presented here seem to point to a 
generic feature that
singularities are not present if $\kappa<0$ whereas singularities do
arise if $\kappa>0$. 

$\bullet$ In the circularly symmetric, static case we have found an
analytical solution for inhomogeneous dust ($p=0$). 
We have also obtained static, circularly symmetric,
numerical solutions for the equation of state $p=\frac{\rho}{2}$
and have briefly analysed the $p=-\rho$ case. 
In the class of circularly symmetric, static solutions found we note that
they are nonsingular for all $\kappa$, except in the $p=0$ case for 
$\kappa>0$. The $p=\frac{\rho}{2}$
solution exhibits the curious feature of being a regular (non-singular)
solution though the radius of the circle $r(l)$ vanishes at 
$l=0$, which usually would imply a singularity. This feature is
exclusively due to the structure of the modified theory (EiBI).

We are aware that in $2+1$ General Relativity
extensive work has been done on cosmological solutions (see for example
\cite{barrow}).
It may be possible to use some of these ideas in $2+1$ EiBI gravity too.
Further, our results for the circularly symmetric, static case 
can easily be generalised to equations of state such as $p=w\rho$ or
the well-known polytropic one. 

In conclusion, we raise a few relevant questions. For instance, one may
ask-- are there black hole solutions in the $2+1$ dimensional EiBI theory?
The answer is surely hidden in the field equations and is worth exploring.
In the context of the cosmological solutions, 
it is important to know about the behaviour of fluctuations about a given
background solution.
A study of such fluctuations can surely be carried out using 
the exact analytical solution 
found here (for a spatially flat FRW line element with $P=0$) as the
background. For the more
general cases (say $P=\frac{\rho}{2}$) 
the numerically obtained spacetimes may be used as backgrounds to
investigate fluctuations, numerically.

Finally, it is possible that our work
on solutions in this toy three dimensional theory 
may provide viable pointers towards finding
new analytical or numerical
geometries in the real four dimensional world.

\end{document}